# PCS — A Roadmap for Exoearth Imaging with the ELT

Markus Kasper[1]
Nelly Cerpa Urra[1]
Prashant Pathak[1]
Markus Bonse[2]
Jalo Nousiainen[3]
Byron Engler[1]
Cédric Taïssir Heritier[1]
Jens Kammerer[1]
Serban Leveratto[1]
Chang Rajani[4]
Paul Bristow[1]
Miska Le Louarn[1]
Pierre-Yves Madec[1]
Stefan Ströbele[1]
Christophe Verinaud[1]
Adrian Glauser[2]
Sascha P. Quanz[2]
Tapio Helin[3]
Christoph Keller[5]
Frans Snik[5]
Anthony Boccaletti[6]
Gaël Chauvin[7]
David Mouillet[7]
Caroline Kulcsár[8]
Henri-François Raynaud[8]

[1] ESO
[2] Institute for Particle Physics and Astrophysics, ETH Zürich, Switzerland
[3] Lappeenranta-Lahti University of Technology, Finland
[4] Department of Computer Science, University of Helsinki, Finland
[5] Leiden University, the Netherlands
[6] LESIA, Observatoire de Paris-Meudon, France
[7] Université Grenoble Alpes, CNRS, IPAG, France
[8] Institut d'Optique, Université Paris-Saclay, France

The Planetary Camera and Spectrograph (PCS) for the Extremely Large Telescope (ELT) will be dedicated to detecting and characterising nearby exoplanets with sizes from sub-Neptune to Earth-size in the neighbourhood of the Sun. This goal is achieved by a combination of eXtreme Adaptive Optics (XAO), coronagraphy and spectroscopy. PCS will allow us not only to take images, but also to look for biosignatures such as molecular oxygen in the exoplanets' atmospheres. This article describes the PCS primary science goals, the instrument concept and the research and development activities that will be carried out over the coming years.

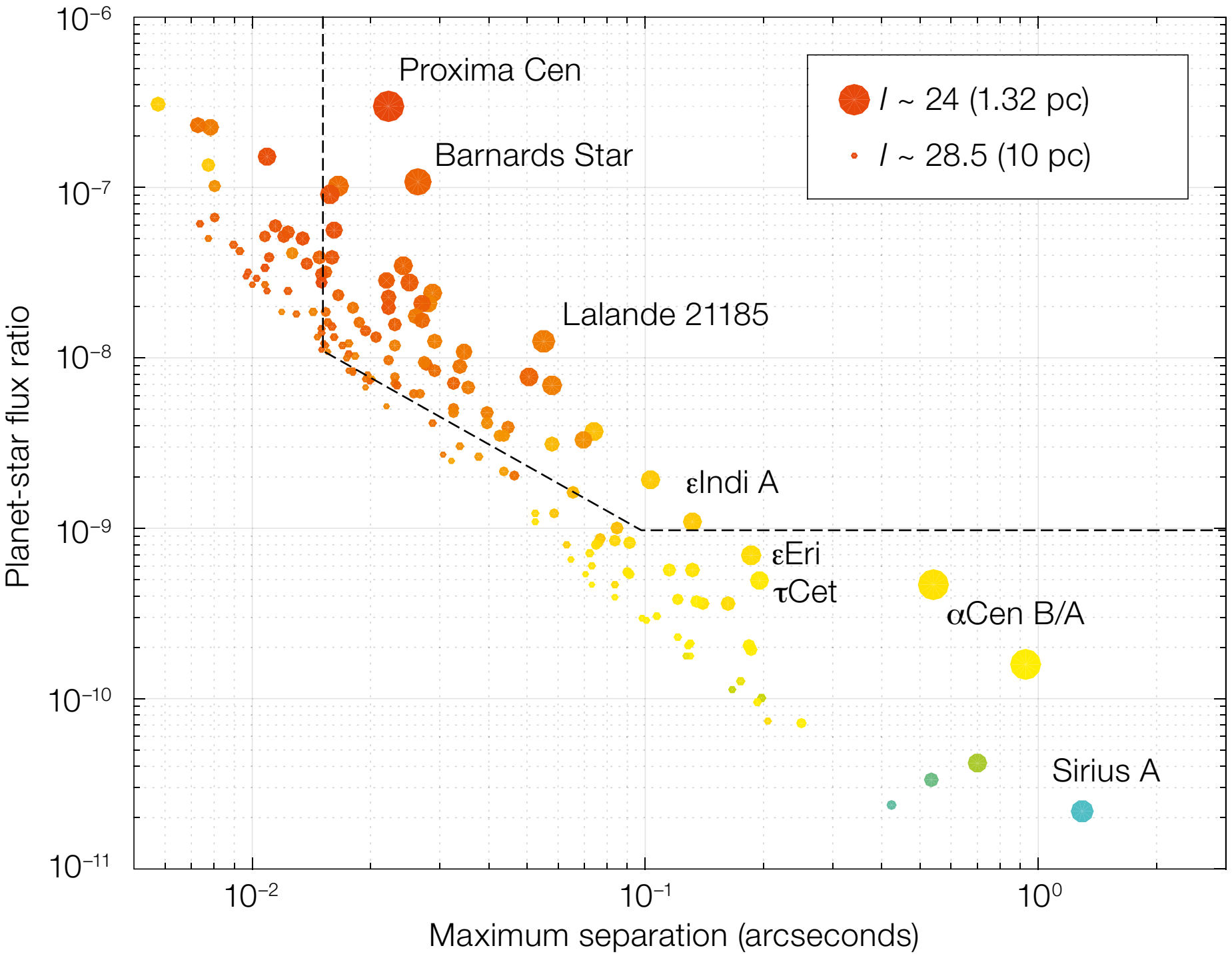

Figure 1. Angular separation and *I*-band flux ratio between hypothetical exoearths (Earth size and insolation, one per star) and parent stars within 10 pc (from the Hipparcos catalogue) observable from Cerro Armazones. The symbol size indicates the planet's apparent brightness, and the colours indicate stellar spectral type (red: M-stars, yellow: solar-type stars). The dotted lines indicate the approximate contrast boundaries for PCS.

## Science Case — nearby exoplanets down to Earth-size

One of the most rapidly developing fields of modern astrophysics is the study of extrasolar planets (exoplanets) and exoplanetary systems. The key goals of the field include understanding the architectures of exoplanetary systems, the formation and evolution of planetary systems, and the composition and structure of exoplanet atmospheres. With over 3000[1] confirmed exoplanets (identified mostly by indirect methods by NASA's Kepler mission), we have developed a basic statistical understanding of the inner regions of planetary systems, i.e., planets with periods less than a few years and orbital separations smaller than a few astronomical units (au), and have thereby made considerable progress towards the first goal. However, the architectures of the outer planetary systems remain essentially unexplored. Given that we do not yet have a complete picture of planetary system architectures, progress towards understanding their formation has been limited. Furthermore, since over 99% of the planets discovered so far have been found indirectly, we have only limited data with which to study and understand the properties of exoplanet atmospheres.

The main reason behind the small number of directly imaged exoplanets is that such observations are extremely challenging. The intensity contrast between the stellar light reflected by an exoplanet and the star itself is less than one part in a million at angular separations of a few tens of milliarcseconds in the case of nearby giant planets discovered by the radial velocity (RV) method, and it becomes even smaller for larger separations between planet and star. Potentially habitable planets with sizes, masses and temperatures like those of Earth are even harder to observe.

Figure 1 shows the approximate *I*-band contrast and angular separation estimated for hypothetical exoearths (planets with Earth-like size and insolation) around the nearest stars.

Therefore in order to directly image and characterise a sizeable number of exoplanets, the combination of telescope and instrument must provide an extremely high contrast and very good sensitivity. Contrast levels around $10^{-8}$ at 15 milliarcseconds and $10^{-9}$ at 100 milliarcseconds are needed for the observation of nearby planetary systems with a limiting exoplanet magnitude of $I$ ~ 27. This would allow us to image giant planets in orbits of a few au already discovered by the RV method, and even to observe potentially habitable planets around very nearby M-stars, as shown in Figure 1. The low-mass M-stars are particularly interesting because around 80% of all stars belong to this group and a considerable number of them are within 20 light-years (~ 6 pc) of the Sun. Temperate small planets have already been found around Proxima Cen (at a distance of 1.3 pc; Anglada-Escudé et al., 2016), Barnard's star (at 1.8 pc; Ribas et al., 2018), Lalande 21185 (at 2.5 pc; Díaz et al., 2019), and Teegarden's star (at 3.8 pc; Zechmeister et al., 2019), and many more are expected to be identified by ongoing and future RV missions (for example, Quirrenbach et al., 2018; Wildi et al., 2017). It is therefore of the utmost scientific importance to understand whether such planets might provide habitable conditions or even show atmospheric fingerprints of biological activity.

The most prominent of these biosignatures is molecular oxygen ($O_2$), which was originally identified as a promising way to find extraterrestrial life in exoplanet atmospheres by Lovelock (1965). It is currently the most easily detectable signal of life in Earth's atmosphere (20% by volume), created as a product of photosynthesis. Most prominent for optical to near-infrared (NIR) observations is the $O_2$ A-band around 765 nm, which consists of a forest of narrow lines. A spectral resolution of several hundred thousand would resolve the unsaturated lines (see López-Morales et al., 2019), and even a spectral resolution of around one hundred thousand would be sufficient to resolve the saturated lines expected to be present in the spectrum of a directly imaged exoearth. In addition, such a high-dispersion spectrum also presents an opportunity to spectrally isolate the planet signal and differentiate it from residual stellar light, thus improving the achievable contrast and sensitivity. Contrast improvements of at least 1:10 000 have been realised with high-dispersion spectroscopy (HDS), which would in principle multiply with the contrast achievable by other methods to supress stellar light at the position of a planet (Snellen et al., 2015).

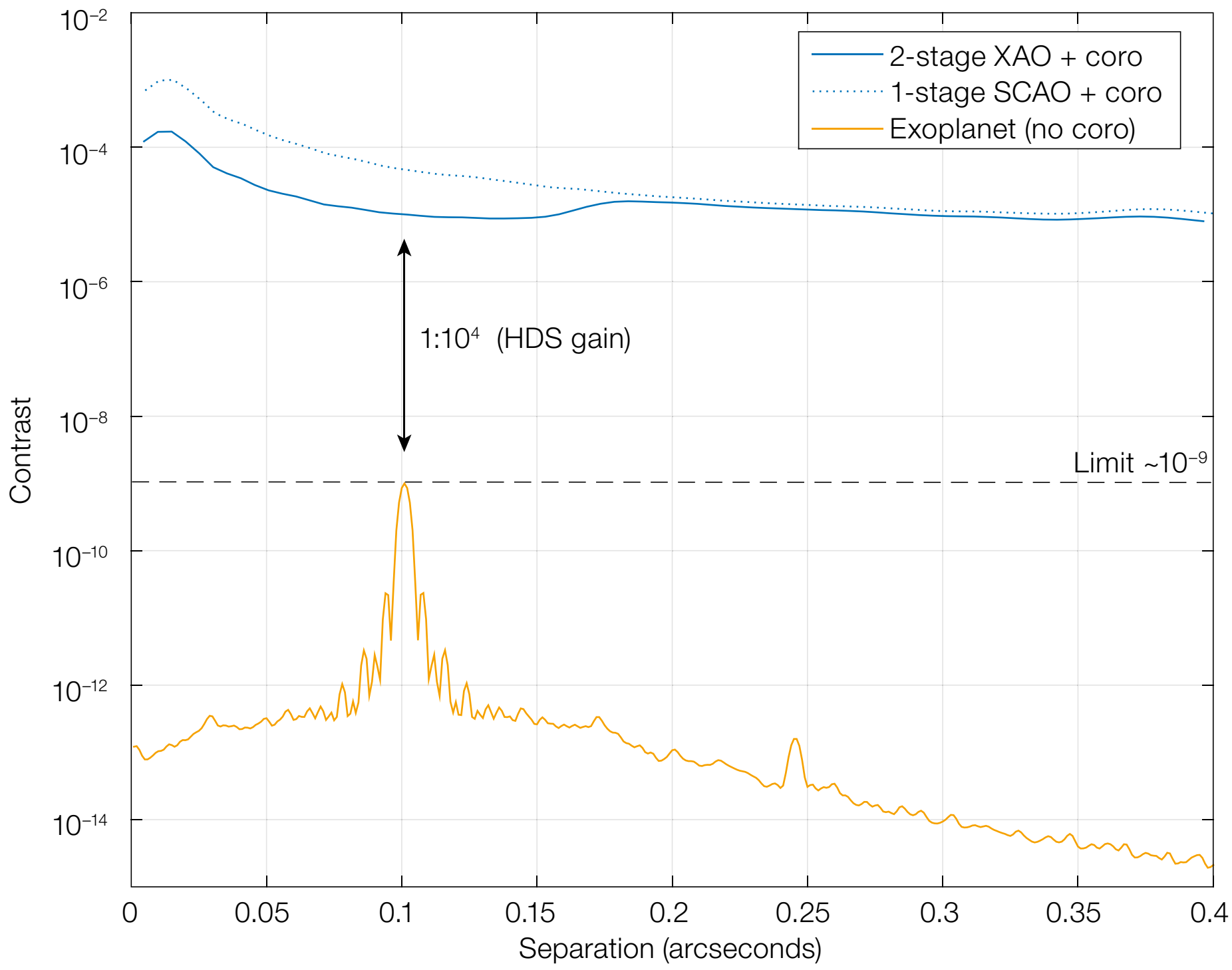

Figure 2. Toy model illustrating how the combination of high-contrast imaging with the ELT and high-dispersion spectroscopy can achieve a contrast of $10^{-9}$ at small angular separations.

Besides the observation of exoearths and the search for biosignatures, a remarkable finding in exoplanetary science in the past decade has been that sub-Neptune planets of around three Earth radii are among the most abundant planets in the solar neighbourhood, despite the absence of any such planet in the Solar System. This planet category sits at the most important transition in the process of planetary formation, namely the onset of runaway accretion of a large gaseous envelope by a rocky core of a critical mass (Pollack et al., 1996). Those planets are so far barely constrained by observations, apart from some $H_2O$ detections in sub-Neptune atmospheres by transit spectroscopy (Tsiaras et al., 2019; Benneke et al., 2019). PCS will detect them in large numbers, allowing us to study their global demographics and explore their role in the conditions of formation of smaller exoearths. PCS will also characterise them and search for molecular features, for example $H_2O$, $CO_2$ and $CH_4$, with medium-resolution spectroscopy in the NIR.

## PCS concept and challenges

To achieve its scientific goals, PCS must provide an imaging contrast of ~ $10^{-8}$ at 15 milliarcseconds angular separation from the star and $10^{-9}$ at 100 milliarcseconds and beyond. In addition, it must provide the spectroscopic capability to observe individual spectral lines due to molecules at optical and NIR wavelengths. The most promising approach for reaching these capabilities is a combination of XAO, coronagraphy and HDS, which must each individually be pushed to the limit.

Figure 2 illustrates this approach. Assuming an HDS contrast of better than $10^{-4}$, the remaining gap to reach the contrast requirements must be provided by the high-contrast imaging (HCI) system consisting of XAO and a coronagraph.

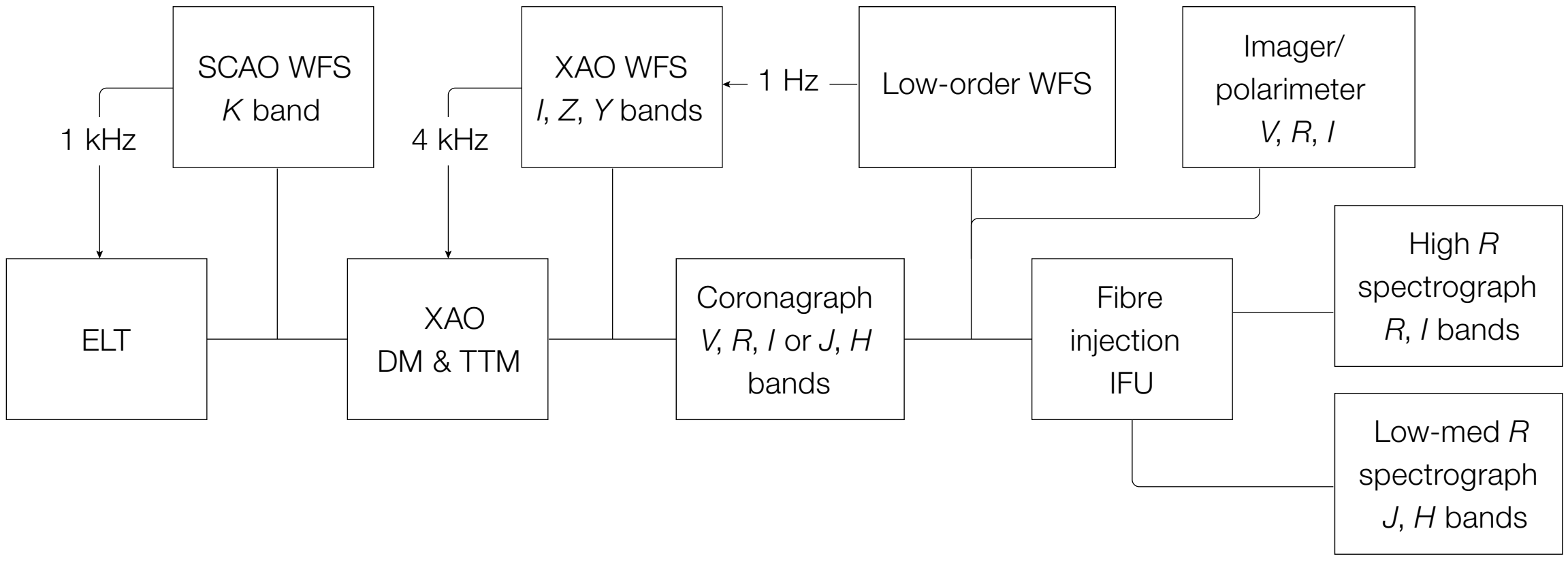

Figure 3. Concept diagram of PCS showing the main building blocks.

The block concept for PCS is shown in Figure 3. The stellar light is suppressed by the HCI as employed by several instruments in operation such as the Spectro-Polarimetric High-contrast Exoplanet REsearch instrument (SPHERE) at ESO's Very Large Telescope (VLT) (Beuzit et al., 2019). The incoming turbulent wavefront will be pre-flattened by a first-stage AO correction with the ELT's M4 mirror (Vernet et al., 2019). The leftover turbulence will be further reduced by a high-speed second stage XAO system working at a wavelength as close as possible to the science wavelength to minimise chromatic residuals. The coronagraph will then strongly attenuate the telescope's diffraction pattern and leave a high-contrast image, as illustrated in Figure 4. The high-contrast raw image is finally fed by an array of single-mode fibres to one of the possible science instruments, for example, a high-dispersion spectrograph working in the red optical *R* and *I* bands where the $O_2$ A band is located.

## Extreme adaptive optics

XAO is a key technology still requiring significant research and development (R&D). It is critically important to minimise the photon noise introduced by stellar light scattered to the position of the nearby planet, which is the main noise source for ground-based exoplanet detection. For example, an XAO-residual halo with contrast on the order of $5 \times 10^{-5}$ at the smallest angular separations forces us to collect $5 \times 10^3$ exoplanet photons for a $1\sigma$ detection of a $10^{-8}$ exoplanet. Correspondingly, a $5\sigma$ detection of this exoplanet would require 25 times more — about $10^5$ — photons. In order to collect that many photons in 10 hours with the ELT (assuming *J* band, 50-nm bandwidth, 10% throughput), the exoplanet must have an apparent magnitude of at most *J* = 26. This is the typical brightness of an exoearth around an M-star at 5 pc. This simple reasoning shows that the required exposure times are inversely proportional to the raw point spread function (PSF) contrast and illustrates how important it is to push XAO performance to its limits.

Figure 4 shows the simulated XAO residual PSF in the *I* band for a VLT-like system. The error budget of the conventional SPHERE-like system shown on the left is dominated by the temporal delay error (see, for example, Guyon, 2005), which shows up as the slightly elongated bright area of increased residuals near the centre. The temporal delay in the system is introduced by wavefront sensor detector integration, detector readout, computation of the correction signal and its application to the deformable mirror (DM). It amounts to at least two update steps of the AO system, during which time the atmospheric turbulence has evolved and no longer perfectly matches the DM correction.

A straightforward approach to reducing the temporal delay is to run the AO system faster. In practice, this is most easily achievable by installing a second small AO system (called the second stage), which corrects a smaller area of the PSF at a much higher update rate, for example 4 kHz instead of the ~ 1 kHz of the first stage. The effect is a greatly improved raw PSF contrast near the star, where most exoplanets are located, as shown in the right panel of Figure 4. This approach is, for example, proposed for the upgrade of SPHERE (called SPHERE+; Boccaletti et al., 2020) and for the potential VLT visitor instrument the high-Resolution Integral-field Spectrograph for the Tomography of Resolved Exoplanets Through Timely Observations (RISTRETTO), aimed at the spectral characterisation of Proxima b (outlined in Lovis et al., 2017).

An even greater performance improvement is expected from predictive control, i.e., by using past measurements to predict the wavefront at the time of the correction. This approach would not only reduce the temporal delay error but could also mitigate the impact of photon noise, thus making the AO system operate better on faint stars (for example, Males & Guyon, 2018). Another important gain could come from controlling the contrast instead of flattening the wavefront, which would require nonlinear control.

A very promising field of research aimed at realising these gains is the application of machine learning techniques. In particular, reinforcement learning (RL) is an active branch of machine learning that provides an automated environment for control. It promises to cope with some of the limitations of current AO systems. Unlike classical control methods, RL methods aim to learn a successful closed-loop control strategy by interacting with the system. Hence, they do not require accurate models of the control loop components and can adapt to a changing environment.

## Fibre-fed spectroscopy

Another field of active R&D for exoplanet detection is the optimised adaptation of high-dispersion integral field spectrographs (IFS) to the HCI case. As HDS also helps to improve the imaging contrast, integral field capability is needed to precisely locate the exoplanets before being able to characterise them. While a fair number of small, nearby exoplanets have already been discovered by the RV method (see above), and many more discoveries are expected for the coming decade, the orbit inclinations and therefore the precise locations of these planets are not accessible by RV.

Fibre-based integral field units may offer some advantages over image slicers or lenslet-array-based solutions. Optical fibres provide a relatively simple means of rearranging the two-dimensional field of view along a single dimension perpendicular to the spectral dispersion direction. In addition, the modal filtering capability of single-mode fibres can be used to create coronagraphs with smaller inner-working angles or higher throughput (Por & Haffert, 2020), and it can also reject random speckles from the XAO system to increase contrast by a factor of a few (Mawet et al. 2017). Single-mode fibres are also well adapted to diffraction-limited imaging, allow light to be efficiently manipulated in photonic integrated circuits, and allow for high-dispersion spectrographs the size of a shoe box. The true potential of such innovative solutions for HCI and HDS, however, remains to be evaluated by laboratory experiments and on-sky observations.

Integral field spectroscopy will be a crucial PCS capability both for improving the contrast performance and for characterising exoplanets and their environments. Current HCI IFS have spectral resolutions of less than 100. To assess and correct instrument aberrations undergoing (chromatic) Fresnel propagation and to detect broad spectral lines and narrow spectral bands, an IFS with a medium spectral resolution (a few thousand) is required for PCS. Since an IFS is always limited by the number of available detector pixels, the medium-dispersion IFS will be able to cover a much larger field and/or a broader spectral range than the high-dispersion IFS. This will be particularly beneficial for detecting and locating exoplanets. The first on-sky testing with a pathfinder has already been carried out by a team led by Leiden University (Haffert et al., 2020). To reach the required technological readiness level for PCS, a visitor instrument could be to be developed and tested with SPHERE at the VLT or the XAO system MagAO-X at the Magellan Clay Telescope (Males et al., 2018).

## PCS planet yield

The scientific capabilities of PCS are best illustrated by the number of small ($R < 4$ Earth radii) exoplanets that PCS would be able to observe as predicted by the Python package P-pop (Kammerer & Quanz, 2018). Assuming planet occurrence rates from NASA's Kepler mission (as in Dressing & Charbonneau, 2015 for M-stars and Kopparapu et al., 2018 for AFGK-stars), circular and randomly oriented orbits, geometric albedos distributed uniformly between 0 and 0.6, and Bond albedos distributed uniformly between 0 and 0.8, we simulated synthetic exoplanets around a sample of 1272 nearby stars within 20 pc of Earth and observable from Cerro Armazones. Then, for each simulated planet, we determined whether it is bright enough to pass the PCS sensitivity limit with an assumed value of 27.5 magnitudes in a 50-hr *I*-band observation, and whether it exceeds the anticipated contrast curve of PCS (dotted lines in Figure 1). Repeating this procedure many times enabled us to statistically assess the number of planets that PCS could detect, as summarised in Figure 5. Of course, PCS will not look at more than a thousand stars for 50 hours each. Our assumption is rather that many of the targets will be identified by ongoing and future RV missions (for example, Quirrenbach et al., 2018; Wildi et al., 2017) and added to the list of already known nearby small exoplanets.

Besides the large number of relatively nearby giant planets already detected by RV observations, PCS has the potential to detect and characterise more than 40 nearby exosuperearths and exoearths. Sub-Neptunes (2–4 Earth radii) are even more likely to be observed, because they are both brighter (since they reflect more light from the host star) and more numerous. Moreover, most detectable planets have equilibrium temperatures of ~ 200–300 K. Colder small planets receive less light from their parent stars and are therefore faint in reflected light, while hotter planets are often too close to their host star to be spatially resolved even by the ELT. Most of the detectable Earth-sized planets orbit M-stars, while a larger number of sub-Neptunes can be observed around AFGK stars.

PCS's high sensitivity to small planets around late-type stars makes it very complementary to thermal-infrared HCI as deployed, for example, in the Mid-infrared

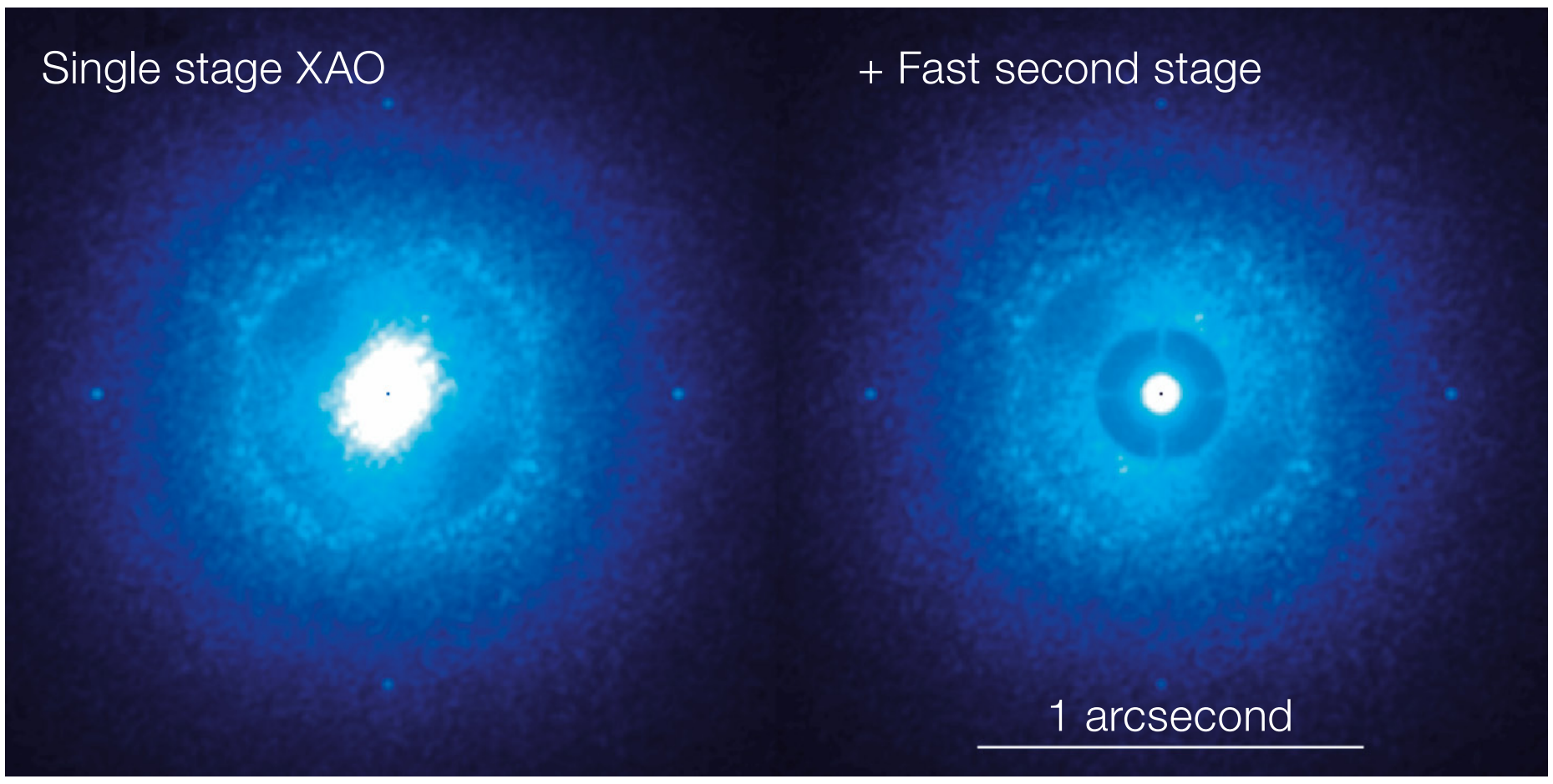

Figure 4. Simulated coronagraphic *I*-band PSFs for an 8-m telescope with a conventional XAO (for example, SPHERE-SAXO) in the left panel, and with an additional fast 4-kHz second stage AO shown in the right panel.

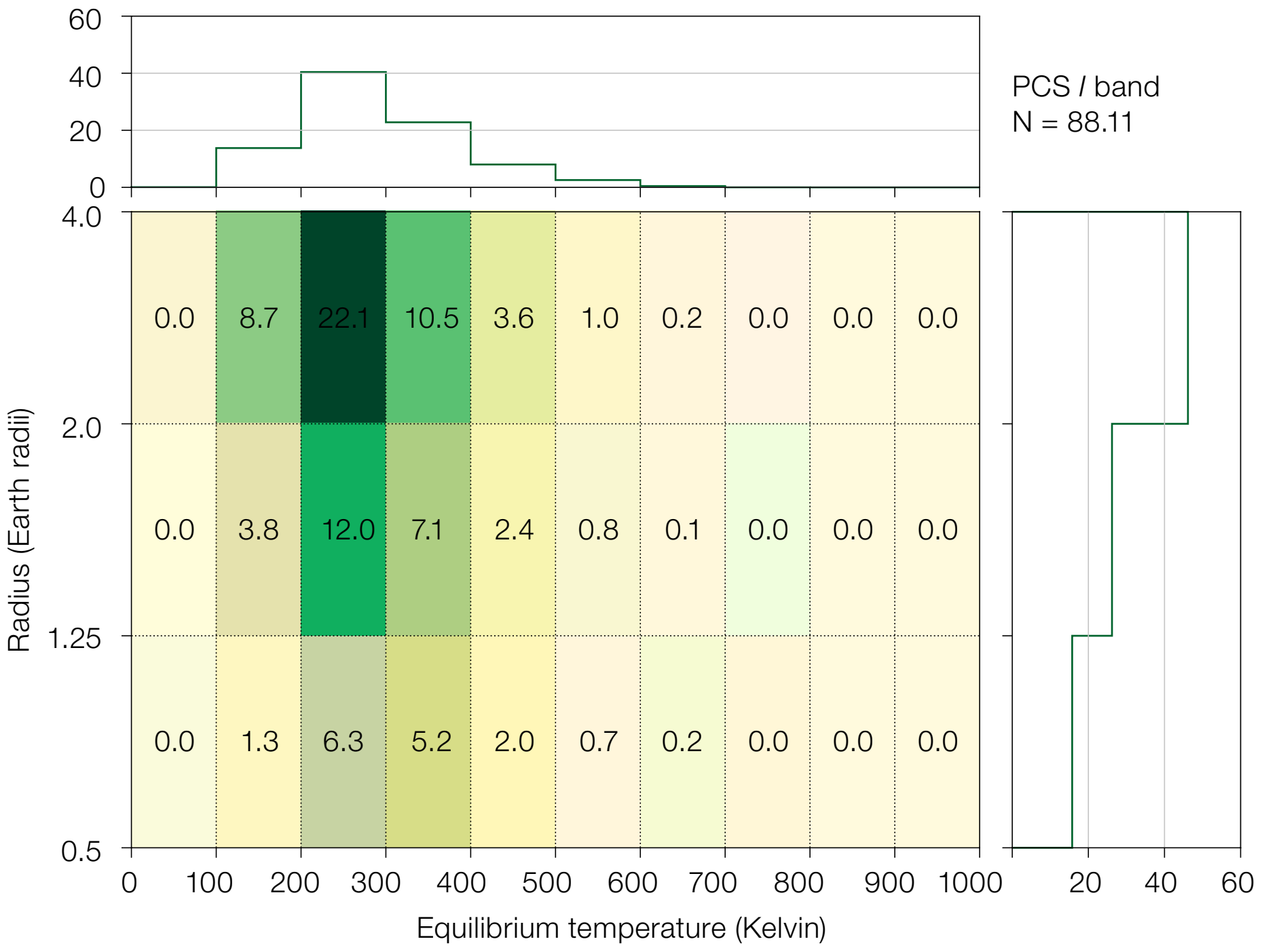

Figure 5. Number of small planet detections in the planet equilibrium temperature vs planet radius plane predicted for PCS observing in the *I* band at quadrature. The marginal histograms show the number of detectable planets as functions of temperature (top) and radius (right).

ELT Imager and Spectrograph (METIS) (Brandl et al., 2018), which is most sensitive to small planets around the handful of very nearby solar-type stars. The visible and NIR also contain many relevant spectral lines and molecular bands, which are crucial for Earth-like exoplanet characterisation and complement the spectral features observable in the thermal-infrared.

## Roadmap towards project start

Prior to the start of the PCS instrument project in a few years, ESO's Technology Development Program is carrying out dedicated R&D activities concentrating on the XAO problem in HCI. In particular, the development of a fast DM with well above 10 000 actuators, excellent positioning resolution (down to 0.06 nm surface deformation) and a small settling time (down to 0.1 ms) is being pursued with European industry. ESO initiated such a development in 2016, and two contracts have been awarded, to ALPAO (France) and to a German consortium comprising Fraunhofer IOF and Physik Instrumente. The goal was to deliver an XAO DM conceptual design backed up by a strong prototyping activity. These two contracts were completed by the end of 2019 and both contractors managed to deliver conceptual designs, based on completely different technologies and compliant with most of the XAO requirements. While the ALPAO design is based on their well-known voice coil approach, Fraunhofer IOF and Physik Instrumente derived a concept based on exchangeable piezoelectric stacked actuators. In each case some limitations have been identified, some of which are being addressed in a second development phase to improve the design and to build a prototype. Other aspects, such as increasing the number of actuators and the corresponding drive electronics, are not yet part of the Technology Development Programme but are projected to be in the roadmap for the next phase of development.

The second major activity, running until 2023, is the development of predictive control methods using classical (see, for example, Kulcsár et al., 2006) and machine learning techniques. Here, the goal is to improve the XAO raw PSF contrast over conventional control methods by factors of 3 to 10 and accordingly reduce the required observing time to obtain a given signal-to-noise ratio by the same factor. Special attention will be given to the feasibility of implementing these methods into the hardware envisaged. We need fast algorithms which have the potential to run in real time at the ELT in about one decade.

The predictive control R&D follows the classical approach. After analysing the problem, promising methods will first be tested in computer models with simulated turbulence and with turbulence measured on-sky by real XAO systems such as, for example, VLT-SPHERE or the Subaru Coronagraphic Extreme Adaptive Optics (SCExAO; Lozi et al., 2018) at the Subaru telescope. This work is being done in collaboration with the ETH Zürich, the Institut d'Optique of the Université Paris-Saclay, the Lappeenranta-Lahti University of Technology, Leiden Observatory and the research organisation TNO in the Netherlands. Furthermore, ESO and the Microsoft Research Laboratory are working to establish a collaboration on Machine Learning-based Predictive Control, which could be of great benefit for PCS R&D.

The best-performing methods will then be implemented and tested in the laboratory. For this, the ETH Zürich and ESO are currently developing an XAO test setup called the GPU-based High-order adaptive OpticS Testbench (GHOST), which will be located in ESO's AO laboratory and will be available in early 2021. GHOST (see Figure 6 for the opto-mechanical setup) will feature a spatial light modulator (SLM) to inject programmable turbulence, measured on-sky by existing XAO instruments. Its XAO system will consist of a Pyramid Wavefront Sensor using a GPU-based real-time computer (RTC) and the freely available CACAO[2] software to control a Boston Micromachines 492-actuator DM. The RTC will be similar to the SCExAO RTC built around the CACAO framework and data format. This concept was also adopted for MagAO-X at the Magellan Clay Telescope and the Keck Planet Imager and Characterizer (KPIC) at the Keck II Telescope (Mawet et al., 2018). Consequently a user community already exists and it will be relatively easy to share code and experience. Interfaces to cameras and DMs frequently used in AO are readily available.

Finally, if a predictive control method shows a significant potential during the GHOST testing period between mid-2021 and the end of 2022, we plan to implement and test it on-sky in 2023. The tests are expected to use the existing SCExAO instrument at the Subaru telescope and possibly a fast second stage AO system added to SPHERE. SCExAO already uses a software architecture similar to that foreseen for GHOST, which should greatly facilitate the porting of algorithms.

In parallel with the R&D on XAO mentioned above, complementary activities are being carried out in the community. The High-Resolution Imaging and Spectroscopy of Exoplanets instrument (HiRISE; Vigan et al., 2018) is a VLT visitor instrument that will use single-mode fibres to feed the high-contrast PSF delivered by SPHERE to the upgraded CRyogenic high-resolution InfraRed Echelle Spectrograph (CRIRES+; Dorn et al., 2016) thereby implementing and validating the general PCS concept. Similar ideas are also being pursued by RISTRETTO, KPIC, MagAO-X and SPHERE+.

After the conclusion of these R&D activities at ESO and within the community, all major ingredients of PCS should be validated and lifted to a high enough level of technology readiness to enable the project to start. PCS will then prepare for the observation of nearby rocky planets and maybe the discovery of an exoplanet that's truly habitable — or even already inhabited — in the 2030s.

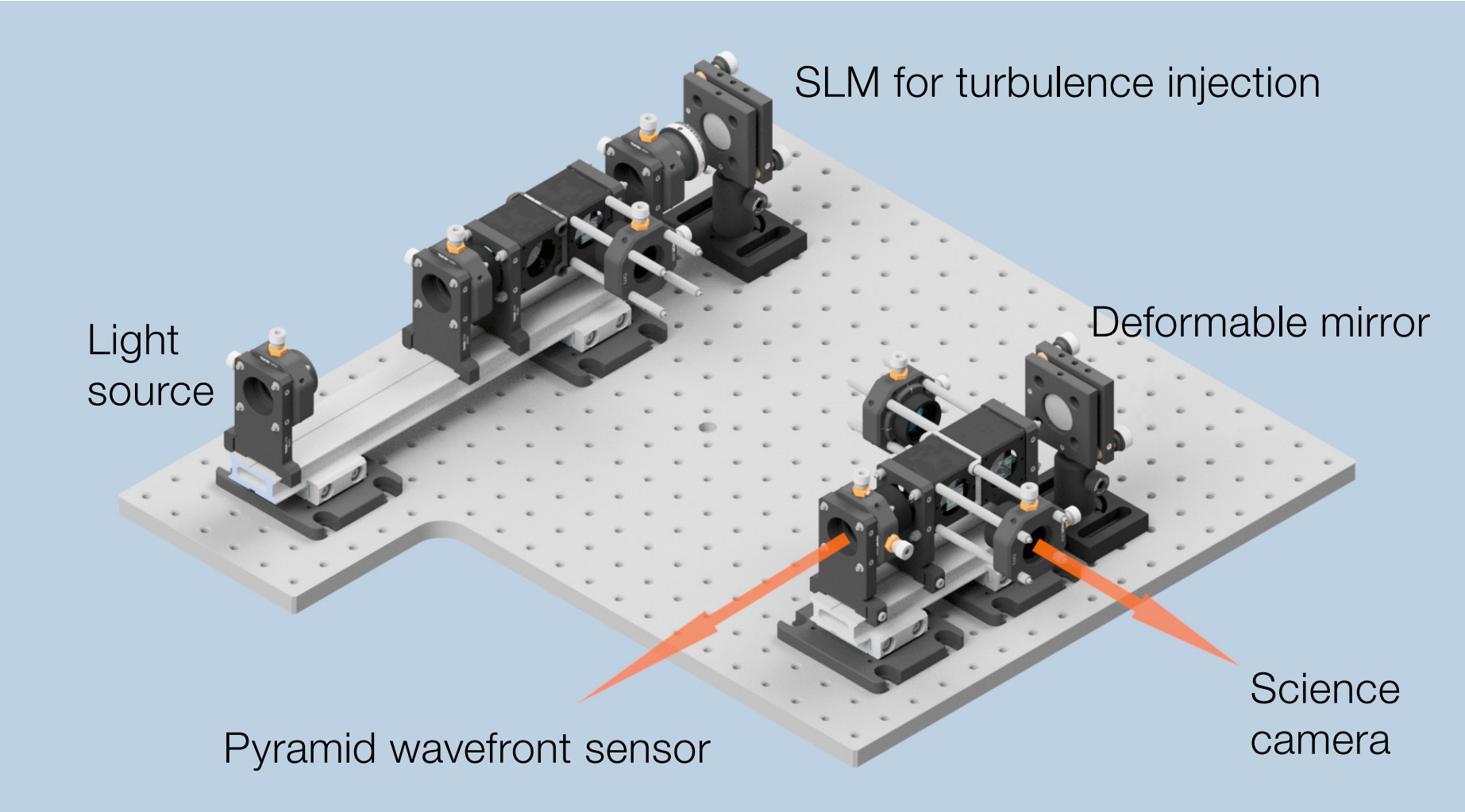

**Figure 6.** Optomechanical setup of GHOST.

## Links

[1] Exoplanet Orbit Database: http://exoplanets.org/
[2] CACAO software: https://github.com/cacao-org/cacao

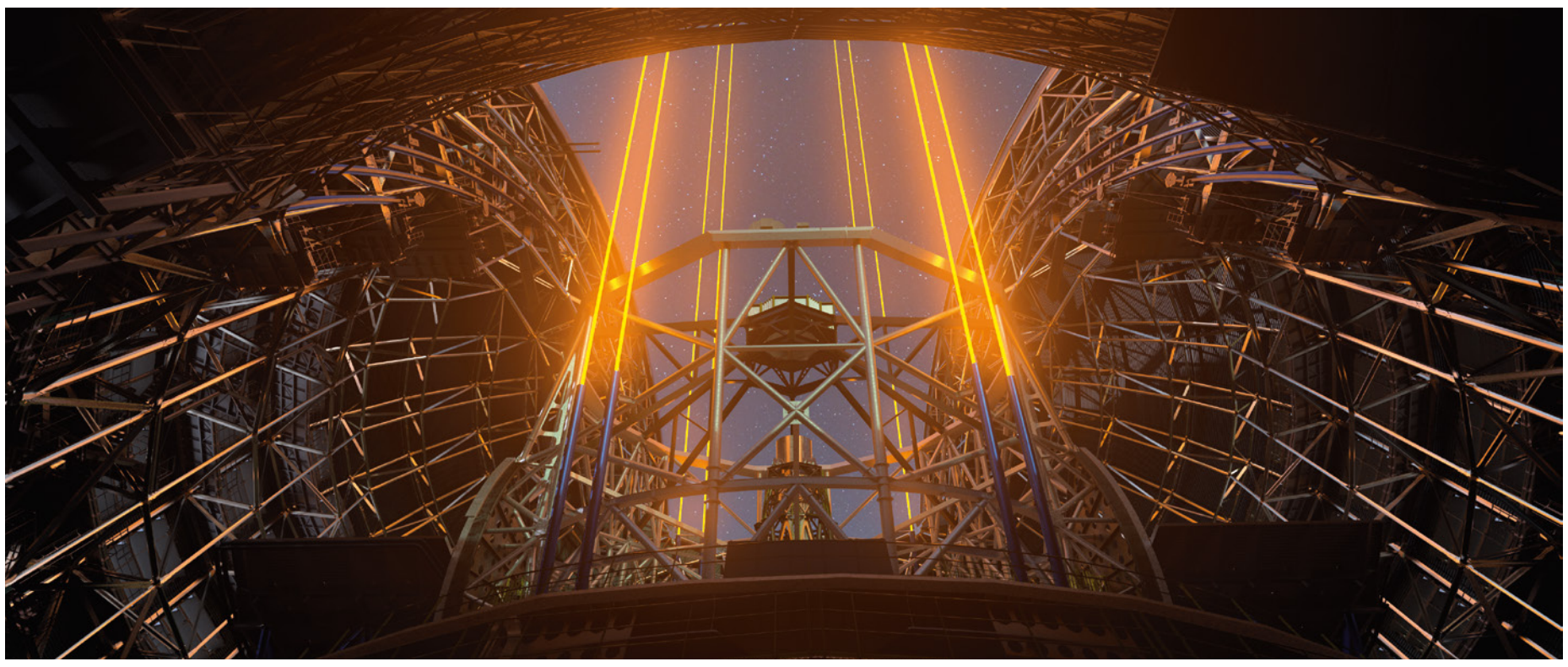

This artistic illustration depicts the laser guide stars of the future Extremely Large Telescope (ELT). Like many other systems on the ELT, the multiple laser guide stars are vital to its operation, helping it adapt to the ever-changing atmospheric conditions above the telescope. This information is sent to the ELT's M4 mirror which will adjust its shape to compensate for the distortion caused by atmospheric turbulence, allowing astronomers to observe finer details of much fainter astronomical objects than would otherwise be possible from the ground.